\documentclass[a4paper,aps,prl,amsmath,amssymb,twocolumn,superscriptaddress,preprintnumbers,noeprint]{revtex4-1} 
\usepackage{amsmath}
\usepackage{verbatim}
\usepackage{graphicx}

\usepackage{color}

\usepackage{xspace}
\usepackage[normalem]{ulem}
 \newcommand{\ket}[1]{\left|#1\right\rangle} 
\renewcommand{\Re}{\text{Re}}
\renewcommand{\Im}{\text{Im}}
\newcommand{\rs}{\rm \scriptscriptstyle}
\newcommand{\MHz}{\text{MHz}\xspace}
\newcommand{\GHz}{\text{GHz}\xspace}
\newcommand{\mum}{\mu \text{m}\xspace}
\def\nn{\nonumber}
\newcommand{\beginsupplement}{        \setcounter{table}{0}
        \renewcommand{\thetable}{S\arabic{table}}        \setcounter{figure}{0}
        \renewcommand{\thefigure}{S\arabic{figure}}     }

\begin{document}
 
\title{Fractional quantum Hall phases of bosons with tunable interactions: From the Laughlin liquid to a fractional Wigner crystal}

\author{Tobias Gra{\ss}}
\affiliation{Joint Quantum Institute, NIST/University of Maryland, College Park, Maryland, 20742, USA}
\author{Przemyslaw Bienias}
\affiliation{Joint Quantum Institute, NIST/University of Maryland, College Park, Maryland, 20742, USA}
\author{Michael J. Gullans}
\affiliation{Department of Physics, Princeton University, Princeton, New Jersey, 08544, USA}
\author{Rex~Lundgren}
\affiliation{Joint Quantum Institute, NIST/University of Maryland, College Park, Maryland, 20742, USA}
\author{Joseph Maciejko}
\affiliation{Department of Physics, University of Alberta, Edmonton, Alberta T6G 2E1, Canada}
\affiliation{Theoretical Physics Institute, University of Alberta, Edmonton, Alberta T6G 2E1, Canada}
\affiliation{Canadian Institute for Advanced Research, Toronto, Ontario M5G 1Z8, Canada}
\author{Alexey V. Gorshkov}
\affiliation{Joint Quantum Institute, NIST/University of Maryland, College Park, Maryland, 20742, USA}
\affiliation{Joint Center for Quantum Information and Computer Science, NIST/University of Maryland, College Park, MD 20742, USA}

\begin{abstract}
Highly tunable platforms for realizing topological phases of matter are emerging from atomic and photonic systems, and offer the prospect of designing interactions between particles. The shape of the potential, besides playing an important role in the competition between different fractional quantum Hall phases, can also trigger the transition to symmetry-broken phases, or even to phases where topological and symmetry-breaking order coexist. Here, we explore the phase diagram of an interacting bosonic model in the lowest Landau level at half-filling as two-body interactions are tuned. Apart from the well-known Laughlin liquid, Wigner crystal phase, stripe, and bubble phases, we also find evidence of a phase that exhibits crystalline order at fractional filling per crystal site. The Laughlin liquid transits into this phase when pairs of bosons strongly repel each other at relative angular momentum $4\hbar$. We show that such interactions can be achieved by dressing ground-state cold atoms with multiple different-parity Rydberg states.
\end{abstract}

\maketitle
\textit{Introduction.}
In a strong magnetic field, a two-dimensional (2D) electron system can form an incompressible liquid phase exhibiting topological order, that is, order without breaking any symmetry \cite{laughlin83}. The Wigner crystal \cite{wigner34,deng16} competes with this liquid phase, and becomes energetically favorable at sufficiently low filling of the lowest ($n=0$) Landau level (LL)  \cite{yoshioka79,lam84}, or via LL mixing \cite{zhao18}. Other phases with broken symmetry have been predicted for half-filled higher LLs ($n>1$), using Hartree-Fock calculations \cite{koulakov96,fogler97,fradkin99} and exact numerical methods \cite{rezayi99,haldane2000}. These phases are characterized by stripe or bubble patterns and have been observed in transport experiments~\cite{lilly99}. Since the early days of fractional quantum Hall (FQH) physics, there have also been different attempts to describe the FQH effect from a crystal phase ansatz \cite{kivelson86,tesanovic89}. Both fractionally quantized and anisotropic transport has been seen experimentally~\cite{xia11,Samkharadze2016}, and different field-theoretic approaches describe this incompressible nematic phase in terms of an effective gauge theory \cite{mulligan10,mulligan11} or assuming the softening of the magnetoroton mode \cite{maciejko13,you14}. Finally, recent numerical work \cite{regnault17} claims evidence of a FQH phase with nematic order in a microscopic model where the third and the fifth pseudopotentials are comparable to the first one.

While electronic materials offer some knobs to control interactions, e.g. via different substrates or patterned metallic gates, their tunability is rather restricted. Thus, it is tempting to study FQH physics, and the interplay of topological order and symmetry breaking, in alternative systems with tunable interactions. Atomic gases are  promising platforms, with the possibility of generating synthetic gauge fields by rotating the system \cite{cooper08} or by optical dressing \cite{goldman14}. More recently, different strategies have also allowed for generating an artificial magnetic field for photons \cite{ozawa18}. These systems are often bosonic, but exhibit similar phases as the electronic systems, including bosonic Laughlin phases and symmetry-broken stripe and bubble phases \cite{cooper05,seki08,grusdt13}. Strikingly, in such controllable systems, a direct transition from the Laughlin liquid to these compressible phases might be achieved, for example by tuning the scattering length via a Feshbach resonance \cite{cooper05}.

Yet a richer phase diagram is expected in the presence of more than one tuning knob. 
In this Letter, 
we study a model with tunable pseudopotentials \cite{regnault17} and consider the bosonic case which is more relevant to atomic, molecular, and optical 
designer quantum Hall systems. 
We focus on a system at filling fraction $\nu=1/2$, 
restricted to the lowest LL, with fixed contact interaction $U_0$, and tunable pseudopotentials $U_2$ and $U_4$ characterizing the scattering strength between bosons with relative angular momentum $2\hbar$ and $4\hbar$. 
Using exact diagonalization, we identify different symmetry-broken phases surrounding the Laughlin liquid. When $U_4>0$ becomes sufficiently large, a new phase with striking features is found: The $N$ bosons form a lattice consisting of $2N$ sites, exhibiting a symmetry-protected two-fold degeneracy of the ground state at zero momentum. 
In contrast to the other symmetry-broken phases, the overlap of the ground state with the Laughlin wavefunction does not drop sharply as the system is tuned from the Laughlin liquid into this new phase. The transition is characterized by a softening of the magnetoroton mode. 
Finally, we demonstrate an experimental proposal based on ground-state atoms dressed with \textit{multiple} Rydberg states, which enable us to explore a wide range of values of pseudopotentials, including the most interesting one with $U_4 \sim U_0$. 

\textit{System.---}We consider a 2D system of $N$ bosons of mass $M$ subjected to a perpendicular gauge field, whose strength is characterized by the cyclotron frequency $\omega_{\rm c}$, or, equivalently by the ``magnetic'' length  $l_B \equiv \sqrt{\hbar/M\omega_{\rm c}}$. The gauge field quenches all bosons into the lowest LL, and interaction between two bosons with relative momentum $\bf q$ (in units $l_B^{-1}$) is described by pseudopotentials~\cite{haldane83}, $U_l = (1/2\pi)^2 \int d{\bf q} \ V_{\bf q} L_{l}(|{\bf q}|^2)e^{-|{\bf q}|^2}$. Here, $V_{\bf q}$ is the potential, and $L_l$ are Laguerre polynomials. In our model, we fix $U_0>0$, tune $U_2$ and $U_4$ from the attractive to the repulsive regime, whereas pseudopotentials with $l>4$ are neglected. 
In the numerics, we consider a rectangular system of size $a \times b$ with periodic boundaries (torus). The number of quantized fluxes $N_\phi$ equals $2N$, that is, $\nu \equiv  N/N_\phi = 1/2$. We choose the gauge potential ${\bf A}$ in the Landau gauge, ${\bf A} \propto (0,x)$, and obtain a single-particle basis of lowest LL wavefunctions $\varphi_j(x,y)$ \cite{yoshioka83}. The quantum number $j$ represents momentum along the $y$-direction.

Evaluating the interaction matrix elements in this basis, we write the Hamiltonian in terms of annihilation/creation operators, $H=\sum_{ijkl} V_{ijkl} a_i^\dagger a_j^\dagger a_k a_l$. The many-body Hilbert space divides into different symmetry sectors: Invariance under magnetic translations leads to conserved (pseudo)momenta $K_x$ \cite{haldane1985}  and $K_y={\rm mod}(\sum_{i=1}^N j_i,N_\phi)$. A sector $K_y$ is connected to $K_y+N_\phi/2$ via a center-of-mass (COM) translation, such that the magnetic Brillouin zone (BZ) can be folded onto a $M\times M$ reciprocal lattice points, with $M$ the greatest common divisor of $N$ and $N_\phi$. Further reduction to the irreducible BZ is possible due to reflection symmetry, leading to an equivalence between $K_{x,y}$ and $-K_{x,y}$ \footnote{An additional $C_4$ symmetry is only present for the special case of a square system, $a/b=1$.}.

\begin{figure}
\centering
\includegraphics[width=0.49\textwidth, angle=0]{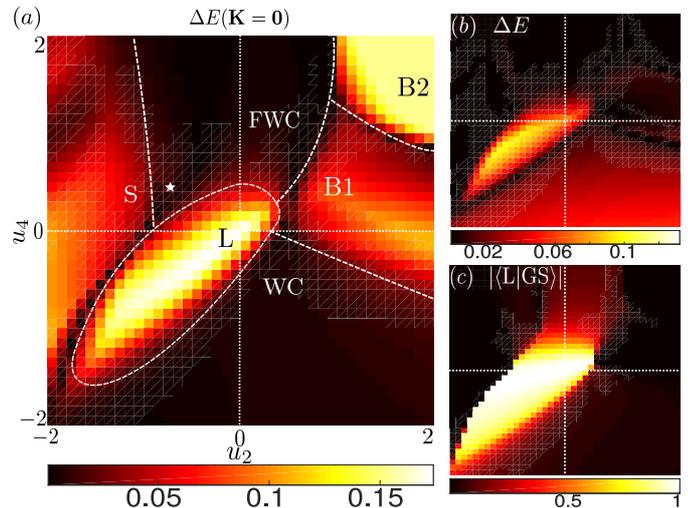}
\caption{\label{phasesN8} $(a)$ Energy gap $\Delta E$ at ${\bf K}=0$ as a function of the pseudopotentials. The infered phase diagram is indicated by white dashed lines. The Laughlin phase (L) is surrounded by a stripe phase (S), bubble phases (B1 and B2), an integer Wigner crystal (WC) phase, and a fractional Wigner crystal (FWC) phase.
The star indicates the point for which the experimental realization is discussed in detail.
 $(b)$ The absolute energy gap versus $u_2$ and $u_4$: Small gaps indicate a symmetry-broken phase, in contrast to the larger gaps seen in the Laughlin liquid phase. $(c)$  The overlap with the Laughlin wavefunction versus $u_2$ and $u_4$: Non-zero overlap persists in the FWC phase. All data was obtained for $8$ bosons on a torus with ratio $a/b=0.9$. 
}
\end{figure}

\begin{figure}
\centering
\includegraphics[width=0.49\textwidth, angle=0]{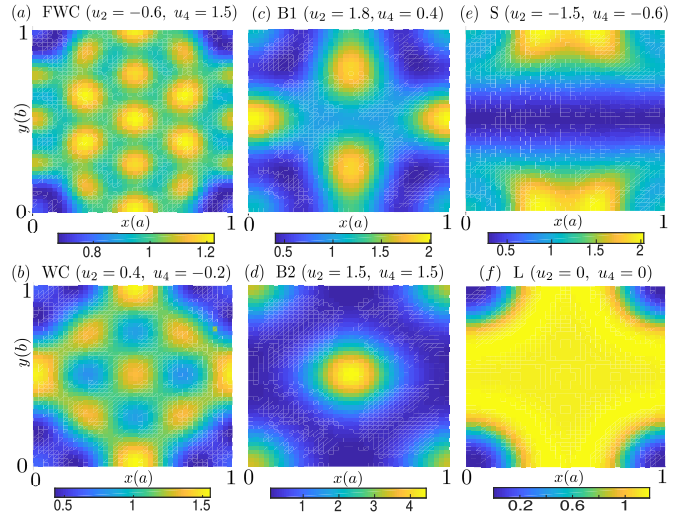}
\caption{\label{grN8} The ground-state pair-correlation function $g_2(x,y)$ for $N=8$ and $a/b$=0.9 for different $u_2$ and $u_4$ corresponding to the different phases: $(a)$ triangular lattice with 1/2 boson per lattice site (fractional Wigner crystal); $(b)$ square lattice with 1 boson per site (Wigner crystal); $(c)$ triangular arrangement of ``bubbles'' (2 bosons per bubble), $(d)$ square arrangement of ``bubbles'' (4 bosons per bubble), $(e)$ clustering along the $x$-axis (stripe), $(f)$ homogeneous Laughlin liquid.
}
\end{figure}

\textit{Numerical Results.---}Via Lanczos diagonalization, we obtained the low-energy eigenstates of $H$ for up to $N=10$, varying the parameters $u_2 \equiv U_2/U_0$ and $u_4 \equiv U_4/U_0$. For $u_2=u_4=0$, the Laughlin state is the unique zero-energy ground state (at ${\bf K}=0$, and, related by COM translation, at ${\bf K}=(0,N_\phi/2)$, where units of $(2\pi/a, 2\pi/b)$ are neglected for notational brevity. By evaluating the energy gaps $\Delta E$ (both the direct gap at ${\bf K}=0$, and the absolute gap) and the overlap with the Laughlin wavefunction, we obtain a putative phase diagram in the $u_2$-$u_4$ space, see Fig.~\ref{phasesN8}. The Laughlin phase is surrounded by phases of broken symmetry with (quasi-)degenerate ground states in different symmetry sectors. A pronounced finite-size gap occurs when $|u_4|$ is large and attractive, see Fig.~\ref{phasesN8}$(b)$.

To identify the order of each phase, we analyze the ground-state pair-correlation function $g_2(z) \propto \big\langle \sum_{i\neq j} \delta(z-z_i+z_j) \big\rangle$, where $z\equiv x+iy$, 
see Fig.~\ref{grN8}. The Laughlin liquid stands out through perfect anticorrelations, $g_2(0)=0$, and a homogeneous particle distribution [Fig.~\ref{grN8}$(f)$]. When $u_2$ is  repulsive and sufficiently strong, clustered lattice configurations become favorable [Fig.~\ref{grN8}(c,d)]. Such ``bubble'' phases are also expected for electronic systems in higher LLs \cite{koulakov96,fogler97,fradkin99,rezayi99,haldane2000}, dipolar gases \cite{cooper05}, and Rydberg systems \cite{grusdt13}. A configuration consisting of a single cluster along one direction appears when $u_2$ is attractive [Fig.~\ref{grN8}$(e)$]. In the case of an attractive $u_4$ potential, we find a square crystal arrangement with one boson per site [Fig.~\ref{grN8}$(b)$]. Of course, compressible phases depend also on the system geometry, chosen as $a/b=0.9$ and $N=8$. Similar results obtained for other system sizes are presented in the 
Supplemental Material \cite{supp}.

\begin{figure}
\centering
\includegraphics[width=0.49\textwidth, angle=0]{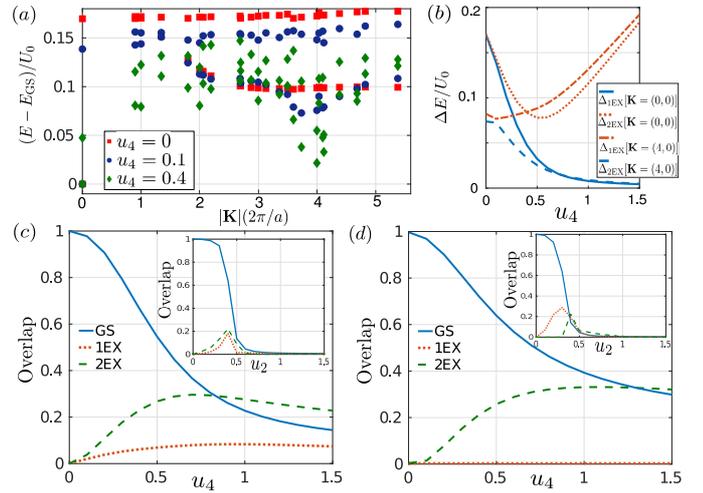}
\caption{\label{trans} $(a)$ Energy of the lowest two eigenstates at each ${\bf K}$, for different $u_4$ (with $u_2=0$). Increasing $u_4$ softens the magnetoroton branch around ${\bf K}=(4,0)$ and ${\bf K}=(2,4)$. $(b)$ Direct energy gaps $\Delta_{1{\rm EX}}$ and $\Delta_{2{\rm EX}}$ of the first and second excited state at ${\bf K}=0$ and ${\bf K}=(4,0)$, as a function of $u_4$ (with $u_2=0$). $(c)$ Overlap between the Laughlin state and the three lowest eigenstates (${\bf K}=0$) of $H$ as a function of $u_4$, with $u_2=0$. The transition into the FWC phase ($u_4\approx0.5$) occurs without a sudden drop of the overlap, in contrast to the transition into a bubble phase shown in the inset (overlap vs. $u_2$, with $u_4=0$). $(b)$ Overlap between the Laughlin magnetoroton state at ${\bf K}=(4,0)$ and the three lowest eigenstates. All data in (a-d) was obtained for $N=8$ and $a/b=0.9$}
\end{figure}

We now turn our attention to the interesting behavior found when $u_4$ is strong and repulsive [Fig.~\ref{grN8}$(a)$]: The pair-correlation function shows a triangular lattice structure with $2N-1$ peaks (plus a deep valley at $z=0$), so we call it a ``fractional Wigner crystal'' (FWC). A half-filled crystal exhibits quantum fluctuations and frustration, and one might speculate that the bosons have fractionalized into semions forming a lattice at filling 1. The transition from the Laughlin liquid into the FWC phase suggests a close relation between the FQH and the FWC phases: As shown in Fig.~\ref{trans}$(a)$, the FWC phase arises through a softening of the magnetoroton mode, a collective excitation branch obtained by a long-wavelength density modulation of the Laughlin state \cite{gmp86}. Finite values of $u_4$ soften this branch near $|{\bf K}|a/(2\pi) \approx 4$, and degeneracy with the ${\bf K}=0$ ground state occurs at $u_4\approx 0.5$, giving rise to a symmetry-broken phase. 
In the same regime, the first-excited state at ${\bf K}=0$ becomes quasi-degenerate, too, and for $u_4 \approx 0.5$ the direct gap to the second-excited state is minimal, see Fig.~\ref{trans}$(b)$. The overlaps of the FWC ground states with the Laughlin wavefunction at ${\bf K}=0$ [see Fig.~\ref{trans}$(c)$], and with the Laughlin magnetoroton state at ${\bf K}=(4,0)$ [see Fig.~\ref{trans}$(d)$] decays smoothly as $u_4$ is increased, but remains finite even deep in the FWC phase.
This behavior is in sharp contrast to the behavior at the boundary between Laughlin and bubble phase, shown in the insets of Fig.~\ref{trans}$(c,d)$: Upon increasing $u_2$ at $u_4=0$, a sudden drop of the overlap to values near zero occurs at the phase boundary. These observations suggest that Laughlin-like behavior remains present in the ground and excited states of the FWC phase.

\begin{figure}
\centering
\includegraphics[width=0.49\textwidth, angle=0]{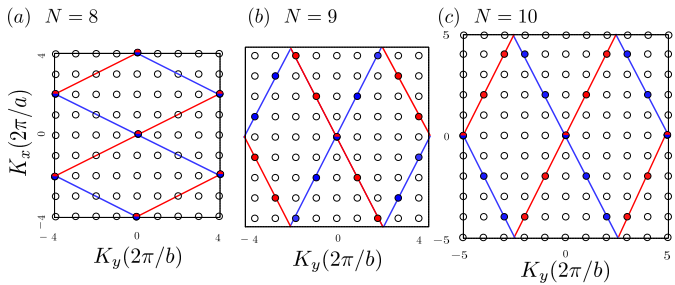}
\caption{\label{BZ} We plot the folded magnetic Brillouin zone for different $N$,
and mark with filled circles the symmetry sectors belonging to the ground-state manifold in the FWC phase. Doubly degenerate sectors are filled with two colors.
The ground states form two stripes (red and blue solid lines), related to each other via reflection, winding twice around the zone.}
\end{figure}

A characteristic feature of each symmetry-broken phase is its tower of states \cite{anderson52,laeuchli16}, that is, the structure of the quantum numbers of the degenerate ground states. This structure reflects the order seen in the pair-correlation function: The triangular bubble phase (B1) [Fig.~\ref{grN8}$(c)$] has degenerate ground states (for $N=8$) at reciprocal lattice vectors ${\bf K}_{mn}=m (2,1) + n(2,-1)$; the square bubble phase (B2) [Fig.~\ref{grN8}$(d)$] at ${\bf K}_{mn}=m (1,1) + n(1,-1)$, and the stripe phase (S) [Fig.~\ref{grN8}$(e)$] at ${\bf K}_{m}=m (0,1)$. In the FWC phase, the ground states form a pair of stripes winding twice around the folded magnetic Brillouin zone, see Fig.~\ref{BZ}. In contrast to conventional stripe phases, the stripes are not parallel to a symmetry axis of the torus, but they are parametrized as ${\bf K}^{i\pm}_m = m (k_x^i,\pm k_y^i)$, with $k_x^i \neq 0 \neq k_y^i$. The pairwise occurrence of these stripes is demanded by reflection symmetry, and leads to characteristic double degeneracies at reciprocal lattice points where the stripes cross. For $N=8$ the two stripes describe exactly the same set of points, and the ground state pattern in reciprocal space matches with a triangular structure, also seen in the correlation function for $N=8$ [Fig.~\ref{grN8}$(a)$]. In contrast, for $N=9$ and $10$, stripe crossings coincide with reciprocal lattice points only at ${\bf K}=(0,0)$. Accordingly, also the correlation function
deviates from the regular lattice structure (see  Supplemental Material \cite{supp}), but still exhibits $2N-1$ peaks.

\begin{figure}[b]
\includegraphics[width=0.49\textwidth, angle=0]{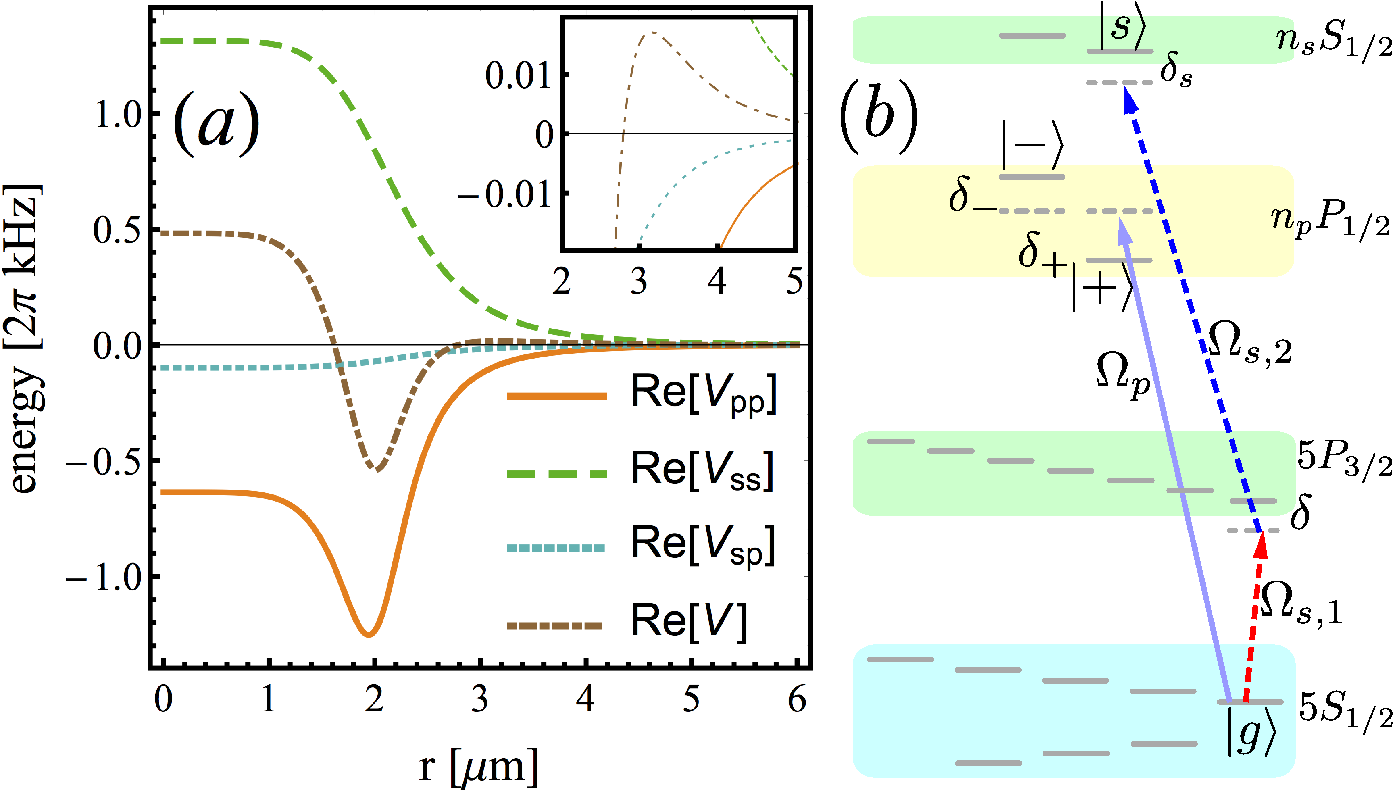}%
\caption{\label{realization} 
Experimental realization. ($a$)  
 The $s-$ and $sp$-dressing lead to a standard soft-core potential (green and blue dashed lines, respectively), whereas $p$-dressing leads to the potential with a sharp dip (orange solid). Together they lead to a hump-dip-hump potential (brown dot-dashed).
The outer hump, relevant for ensuring that $U_4>0$, is shown in the inset.
($b$)
Level scheme for the dressing of the ground state $\ket{g}$ with Rydberg states $\ket{s}$ and $\ket{+}$. 
}
\end{figure}

\textit{Experimental realization.---} The phases discussed above can be realized using cold ground-state alkaline 
atoms dressed with Rydberg states~\cite{Saffman2010,Low2012} in a synthetic magnetic field  generated by rotating the system \cite{cooper08}.
Typical interactions between $s$-state-Rydberg-dressed atoms (dashed green curve in Fig.~\ref{realization}$(a)$) saturate for distances smaller than the so-called Rydberg blockade radius, a phenomenon studied in the context of FQH states in Ref.~\cite{grusdt13}. 
For dressing with Rydberg $p$ states, the interaction can be non-monotonous  (solid orange curve in  Fig.~\ref{realization}$(a)$) as a function of distance~\cite{VanBijnen2015,Glaetzle2015}.
Each of these two cases enable us to explore part of the phase diagram;
however, in order to access the FWC regime we need even more flexibility in the shape of interactions.
We propose to combine the $s$- and $p$-state dressing to achieve a hump-dip-hump potential  (dot-dashed curve in Fig.~\ref{realization}$(a)$). 

As an example  (see Fig.~\ref{realization}$(b)$) we consider ground-state $\ket{g} = \ket{F=2,m_F=2}$  atoms of $^{87}$Rb weakly dressed with two Rydberg states: (i) an $n_p P_{1/2}$  state $\ket{+}=\ket{n_p,m_J=1/2}$ using a laser field with Rabi frequency $2\Omega_p$ and detuning  $\delta_{+}$, and (ii) an $n_s S_{1/2}$  state $\ket{s}=\ket{n_s,m_S=1/2}$ using an effective Rabi frequency $2 \Omega_s$ and detuning $\delta_s$. The coupling to $\ket{s}$ is achieved using a two-photon transition with single-photon detuning $\delta$ and two Rabi frequencies $2 \Omega_{s,1},2 \Omega_{s,2}\ll |\delta|$, leading to $\Omega_{s}=-\Omega_{s,1}\Omega_{s,2}/\delta$.
Without interactions and for $|\frac{\Omega_{s/p}}{\delta_{s/p}}|\ll 1$, the dressed state takes the form $\ket{d}=\ket{g}-\frac{\Omega_{s}}{\delta_{s}}\ket{s}-\frac{\Omega_{p}}{\delta_{p}}\ket{+}$.
For weak dressing, the total two-body interaction $V(r)$ between two $\ket{d}$ states is a sum over separately calculated potentials, $V=V_{pp}+V_{ss}+2V_{sp}$, where $V_{a a'}$ arises due to the interaction of Rydberg states $a$ and $a'$. 

By choosing $|n_p-n_s|\gg 1$, we can neglect direct dipolar coupling between two-atom states $\ket{s\pm}$ and $\ket{\pm s}$ (where $\ket{-} = \ket{n_p,m_J=-1/2}$) and describe the interactions using only diagonal 
van der Waals (vdW) potentials. 
The vdW interaction between $\ket{\pm}$, assuming a magnetic field 
 perpendicular to the 2D plane, is
\begin{equation}
\frac{1}{r^6}\left(
\begin{array}{cccc}
 \alpha -\beta  & 0 & 0 & \beta  \\
 0 & \alpha +\frac{\beta }{3} & -\frac{\beta }{3} & 0 \\
 0 & -\frac{\beta }{3} & \alpha +\frac{\beta }{3} & 0 \\
 \beta  & 0 & 0 & \alpha -\beta  \\
\end{array}
\right)
\end{equation}
in the $\{++,+-,-+,-- \}$ basis,
with $\alpha/2\pi=690.2\, \MHz\,\mum^6$ and $\beta/2\pi=6204.3\, \MHz\,\mum^6$ for $n_p=62$.
This leads to the effective interaction $V_{pp}$ between the $p$ components of the dressed state, which, within a fourth-order-perturbation calculation, equals ($\hbar=1$)
\begin{equation}
\frac{2 \Omega _p^4 \left[\alpha  (\alpha -2 \beta )+2 \delta _- r^6 (\alpha -\beta )\right]}{\delta _+^3 \left\{\alpha  (\alpha -2 \beta )+2 r^6 \left[\delta _+ (\alpha -\beta )+\delta _- \left(\alpha -\beta +2 \delta _+ r^6\right)\right\}\right]}.\nn
\end{equation}

The interactions $V_{ss}$ and $V_{sp}$ arise from standard dressing of each ground-state atom with a single Rydberg level interacting via a vdW $C_6/r^6$ potential. In this case, the interaction takes the form 
\begin{equation}
\frac{C_6 \left(\delta _1+\delta _2\right) \Omega _1^2 \Omega _2^2}{\delta _1^2 \delta _2^2 \left[C_6+\left(\delta _1+\delta _2\right) r^6\right]},
\end{equation}
where, for $V_{ss}$, we set $\delta_1=\delta_2 = \delta_s$,  $\Omega_1=\Omega_2 = \Omega_s$, and $C_6=C_{ss}$, while, for $V_{sp}$, we set $\delta_1=\delta_s$, $\delta_2=\delta_+$, $\Omega_1=\Omega_s$, $\Omega_2=\Omega_p$, and $C_6=C_{sp}$.

The strength of $V_{ss}$ and $V_{sp}$ relative to $V_{pp}$ can be tuned via Rabi frequencies, detunings, and principal quantum number $n_s$.
By setting $\delta_+/2\pi=-21.41\,\MHz$, $\delta_-/2\pi=16.15\,\MHz$, we achieve a resonance-free hump-dip $V_{pp}$ potential~\cite{Glaetzle2015}, and by choosing $\delta_s/2\pi=17.28\,\MHz$ and $n_s=52$, such that  $C_{sp}/(2\pi\,\MHz\,\mum^6)= -682.08<0$ and $C_{ss}/(2\pi\,\MHz\,\mum^6)=3918.89>0$, the other two potentials are also resonance-free, and $V_{sp}$ is much weaker than $V_{ss}$, with Rydberg blockade radius between $s$ states $r_{ss}=2.2\mum$~\footnote{
In our case, the hyperfine splitting in the $n_p P$ manifold is $\Delta_{\rs hfs,n_p}\approx 0.4\,\GHz$, which means that 
the corrections to the vdW potential can be significant. Nevertheless, we expect that our results should hold qualitatively and, moreover, by working with smaller $\delta_\pm$ and $\delta_s$ we can ensure that the corrections to the vdW potential will be weaker. 
}.
We choose $l_B=2\mum\sim r_{ss}$, so that we can neglect $U_j$ for $j>4$.
By choosing optimal Rabi frequencies, we can still operate in a weakly dressed regime and simultaneously access the hardest to achieve regime of strong $U_4$, corresponding to the hump-dip-hump potential in Fig.~\ref{realization}$(a)$. 
Specifically, for $\Omega_p/(2\pi)=1.32933\,\MHz$ and $\Omega_s/(2\pi)=1.35679\,\MHz$,  we get $\{u_2,u_4,u_6\} = \{-0.72,0.45,0.11\}$ with $U_0/2\pi=0.00061\,$kHz and $\Im[U_l]/\Re[U_l]<0.01$.
Finally, if the LL gap  $\omega_{\rm c}$  is larger than the intra-LL pseudopotentials and the 
relevant inter-LL interactions   (both are $\sim U_0$), we can neglect higher LLs. For $l_B=2\mum$, this is indeed the case,  $U_0\ll \omega_{\rm c} =2\pi\times 0.029\,$kHz.
By changing the detunings and Rabi frequencies and possibly varying them in time, one can investigate other phases of the phase-diagram in Fig.~\ref{phasesN8} and study transitions between them.

\textit{Summary.---}We studied a FQH system that exhibits a density modulation when higher pseudopotentials (in particular $U_4$) are on the order of $U_0$ and proposed realizing such exotic interactions with Rydberg-dressed atoms. Our scheme allows us to explore FQH scenarios beyond those found in electronic systems with Coulomb interactions. Our findings point towards an interplay of two fundamental concepts, topological order and symmetry breaking, which both appear to be present in our system.
%

\begin{acknowledgments}
We acknowledge discussions with R.\ Belyansky, A.\ Gromov, and S.\ Sondhi. T.G. acknowledges funding by the NSF through the PFC@JQI. R.L. is supported by a NIST NRC Research Postdoctoral Associateship Award. P.B., R.L., and A.V.G.~acknowledge funding by AFOSR, NSF PFC at JQI, ARO, ARO MURI, ARL CDQI, NSF Ideas Lab, Department of Energy ASCR Quantum Testbed Pathfinder program, and Department of Energy BES Materials and Chemical Sciences Research for Quantum Information Science program. J. M. was supported by NSERC grant $\#$RGPIN-2014-4608, the CRC Program, CIFAR, and the University of Alberta. The authors acknowledge the University of Maryland supercomputing resources (\url{http://hpcc.umd.edu}) made available for conducting the research reported in this paper.
\end{acknowledgments}


%



\clearpage
\section*{Supplemental material}

\beginsupplement
\setcounter{equation}{0}
\renewcommand{\theequation}{S\arabic{equation}}

This Supplemental Material consists of two Sections. Sec. I extends the numerical study of the main text to other system sizes and torus ratios. Sec. II provides characteristic energy spectra of each phase. 

\section{Dependence of numerical results on system size and/or geometry}
Most of our numerical data presented in the main text was for $N=8$ bosons, on a torus with axes ratio $a/b=0.9$. For incompressible phases, system size and geometry tend to play only a minor role. This is no longer true when the system becomes compressible. In this section of the Supplemental Material, we present some additional data obtained by varying the axis ratio of the torus, and by considering systems of $N=9$ and $N=10$ bosons.

\subsection{Phase diagram}
\begin{figure}
\centering
\includegraphics[width=0.49\textwidth, angle=0]{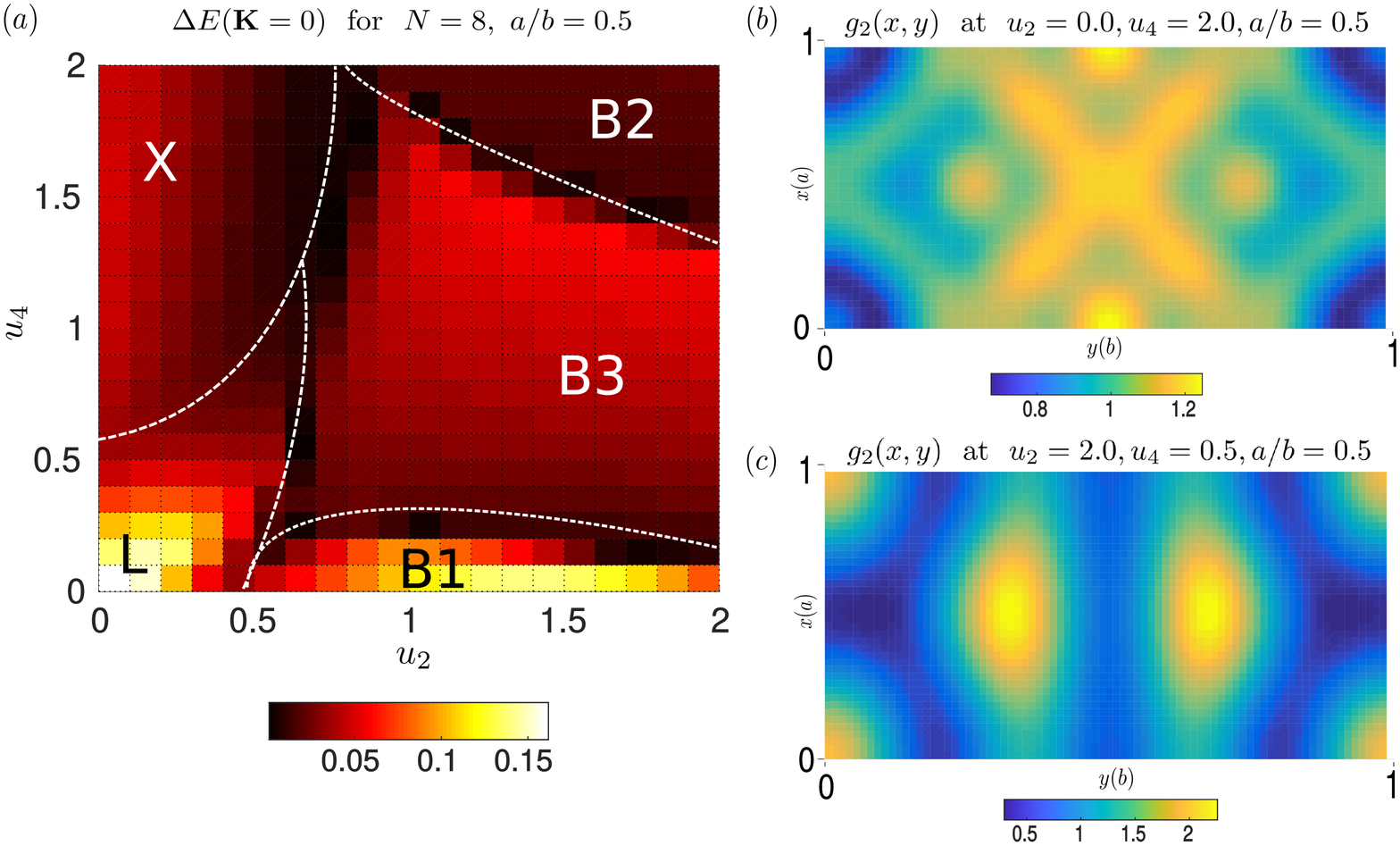}
\caption{\label{squeezedphases} (a) By plotting the direct energy gap at ${\bf K}=0$, we draw a phase diagram in the parameter space given by repulsive $u_2$ and $u_4$. As in the main text, our data was obtained for $N=8$ bosons, but now the torus is squeezed to an axis ratio $a/b=0.5$. As for the less anisotropic case shown in the main text, this squeezed torus also hosts the Laughlin phase (L), and the bubble phases B1 and B2, but now a third bubble phase B3 lies in between those two regimes. The FWC crystal has been deformed,
showed in the pair-correlation function in panel (b). We thus denote this phase by an X. Panel (c) shows the pair-correlation function of the B3 phase, consisting of three bubbles.
}
\end{figure}

First, we consider the phase diagram on a strongly squeezed torus with axis ratio $a/b=0.5$.  Here, we restricted our study to fully repulsive interactions, $u_2,u_4>0$. Again the direct gap at ${\bf K}=(0,0)$ serves as an indicator of phase boundaries. We plot the gap for $N=8$ in Fig. \ref{squeezedphases} (a). The data now suggests the existence of an additional phase. By evaluating the ground-state pair correlation functions, we identify the different phases. As before, we find the Laughlin liquid, and the bubble phases B1 and B2, which are now separated by an intermediate phase B3 consisting of three stripes, see Fig. \ref{squeezedphases} (c). Moreover, it turns out that the FWC is strongly deformed, and does not exhibit $N_\phi$ separate peaks anymore, see Fig. \ref{squeezedphases} (b). In terms of overlap, this region is still smoothly connected to the Laughlin state.

\begin{figure}
\centering
\includegraphics[width=0.49\textwidth, angle=0]{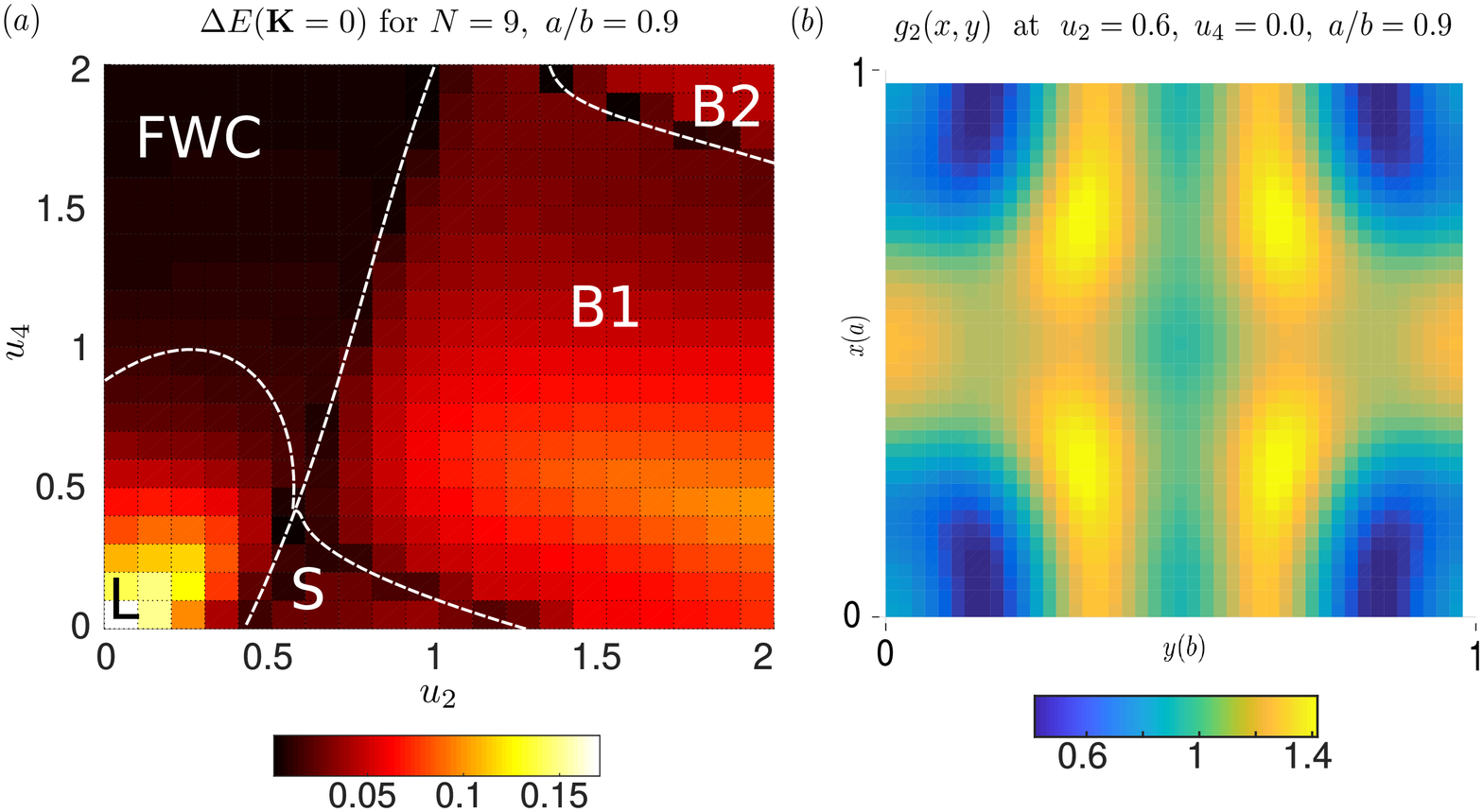}
\caption{\label{N9phases} (a) By plotting the direct energy gap at ${\bf K}=0$, we draw a phase diagram in the parameter space given by repulsive $u_2$ and $u_4$. Here we use the same axis ratio as in the main text ($a/b=0.9$), but increase the system size to $N=9$ bosons and $N_\phi=18$ fluxes. We find a similar phase diagram, hosting the Laughlin phase (L), the fractional Wigner crystal phase (FWC), the bubble phases B1 and B2, plus an additional phase S at intermediate values of $u_2$. (b) By evaluating the ground state pair-correlation function, we identify the intermediate phase S as a phase consisting of three stripes along the $x$-axis.
}
\end{figure}

Next, we compare the phase diagram obtained for $N=8$ bosons with the one for $N=9$, while keeping the geometry unchanged ($a/b=0.9$). We find all the phases which were present also for $N=8$ (Laughlin liquid, FWC, bubbles B1 and B2), plus an additional phase consisting of three stripes. The pair-correlation function for this phase is shown Fig.~\ref{N9phases} (b). In the phase diagram of Fig. \ref{N9phases} (a), the stripe phase occurs at intermediate values of $u_2$, and small values of $u_4$. Such a phase has also been seen before in a system of dipolar atoms \cite{cooper05}. Apart from this intermediate stripe phase, the phase diagrams at $N=8$ and $N=9$ are qualitatively the same.

\begin{figure}
\centering
\includegraphics[width=0.49\textwidth, angle=0]{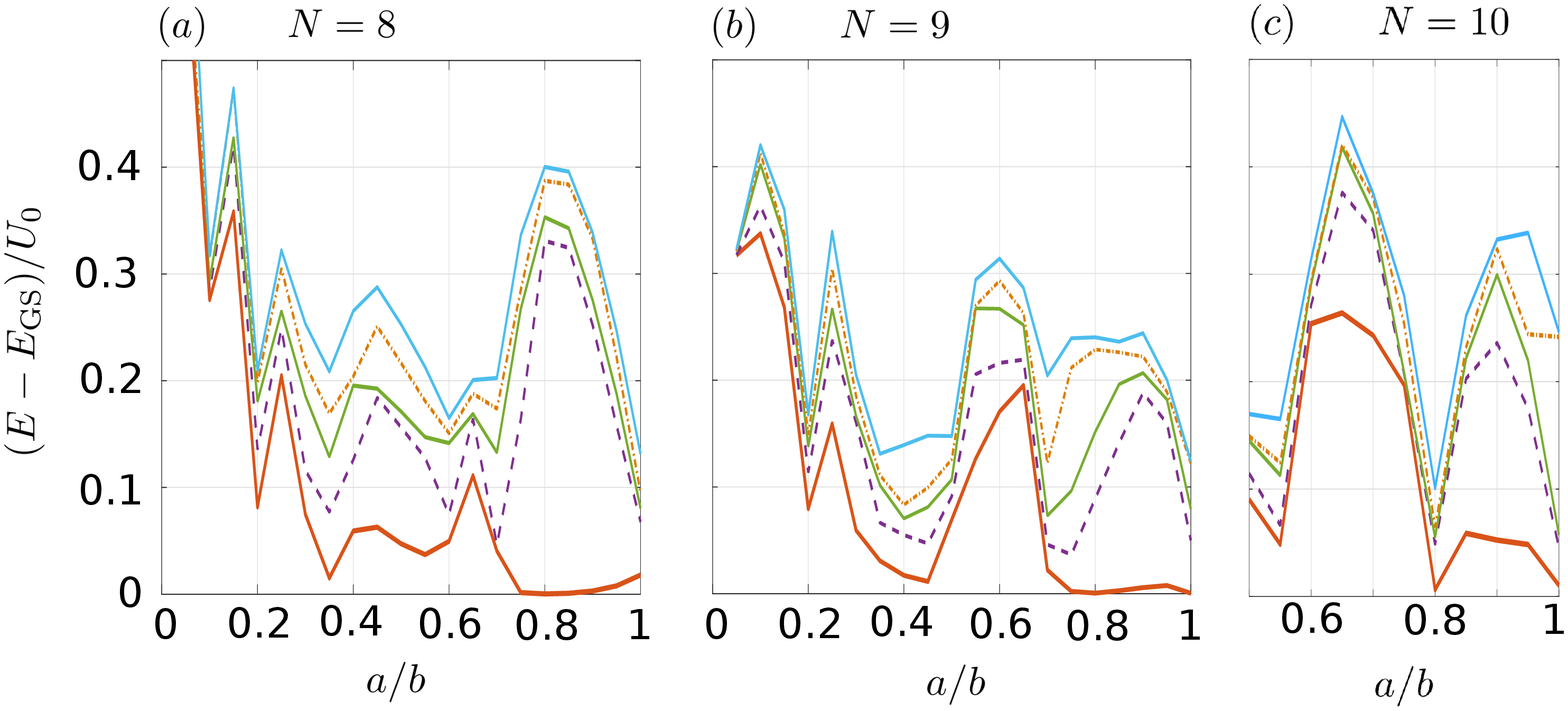}
\caption{\label{squeezing} For different system sizes $N=8,9,10$, we plot the energies of the five lowest excitations at ${\bf K}=0$, relative to the ground state energy, as a function of the axis ratio $a/b$. Thus, the red lines correspond to the direct gap which, due to the two-fold degeneracy, vanishes in the FWC phase. The degenerate regime extends from the isotropic torus ($a/b=1$), to values about $a/b=0.8$. All data was obtained at $u_2=0$ and $u_4=2.0$.
}
\end{figure}

\begin{figure}
\centering
\includegraphics[width=0.49\textwidth, angle=0]{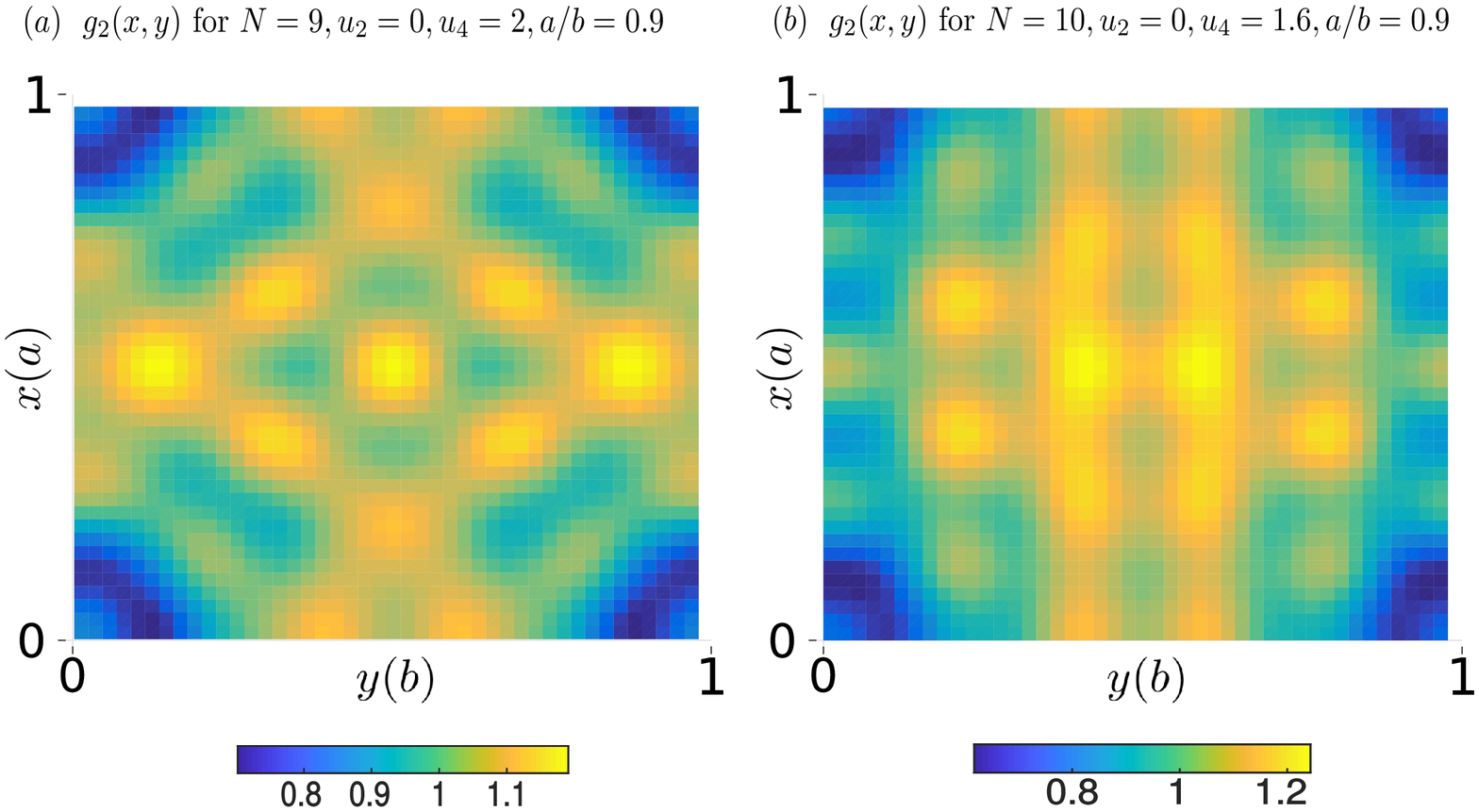}
\caption{\label{N910FWC} We plot, for $N=9$ (a) and $N=10$ (b), the ground state pair-correlation function in the FWC phase. The number of peaks (+origin) is equal to $2N$.
}
\end{figure}

\subsection{Fractional Wigner crystal}
The symmetry-breaking in the FWC phase has a characteristic two-fold ground state degeneracy at ${\bf K}=0$. Thus, to study the robustness of this phase upon changes in the geometry, we may just track the lowest eigenenergies at  ${\bf K}=0$ as a function of the axis ratio $a/b$. This data, for $N=8,9,10$ is shown in Fig. \ref{squeezing}. We note that, even in the FWC phase, the ground state degeneracy is slightly lifted at $N=10$, but this lifting is sufficiently small to identify a quasi-degenerate regime. With this, the characteristic degeneracy is found to extend from the isotropic limit $a/b=1$ down to squeezed ratios $a/b \approx 0.8$.

For larger deviations from the isotropic case, the correlation function does not exhibit $N_\phi$ separate peaks anymore, see Fig. \ref{squeezedphases}(b). The question arises whether such a geometry dependence would also occur in a large system on a flat plane? Due to the non-liquid behavior of the phase, finite-size effects play an important role for choosing the energetically most favorable arrangement. Squeezing the torus enhances finite-size effects, as one axis becomes very short. We therefore believe that the phases obtained near $a/b=1$ are most likely the ones which will be seen in a thermodynamically large system.

By plotting the pair-correlation function of the ground state within the two-fold degenerate regime, we verify that the state exhibits fractional crystal order for different system sizes ($N=9$ and $N=10$). As seen in Fig. \ref{N910FWC}, the number of peaks plus the origin (i.e. the position of the probe particle), is given by $2N$, that is, we always get a half-filled lattice. 

\begin{figure*}
\centering
\includegraphics[width=0.9\textwidth, angle=0]{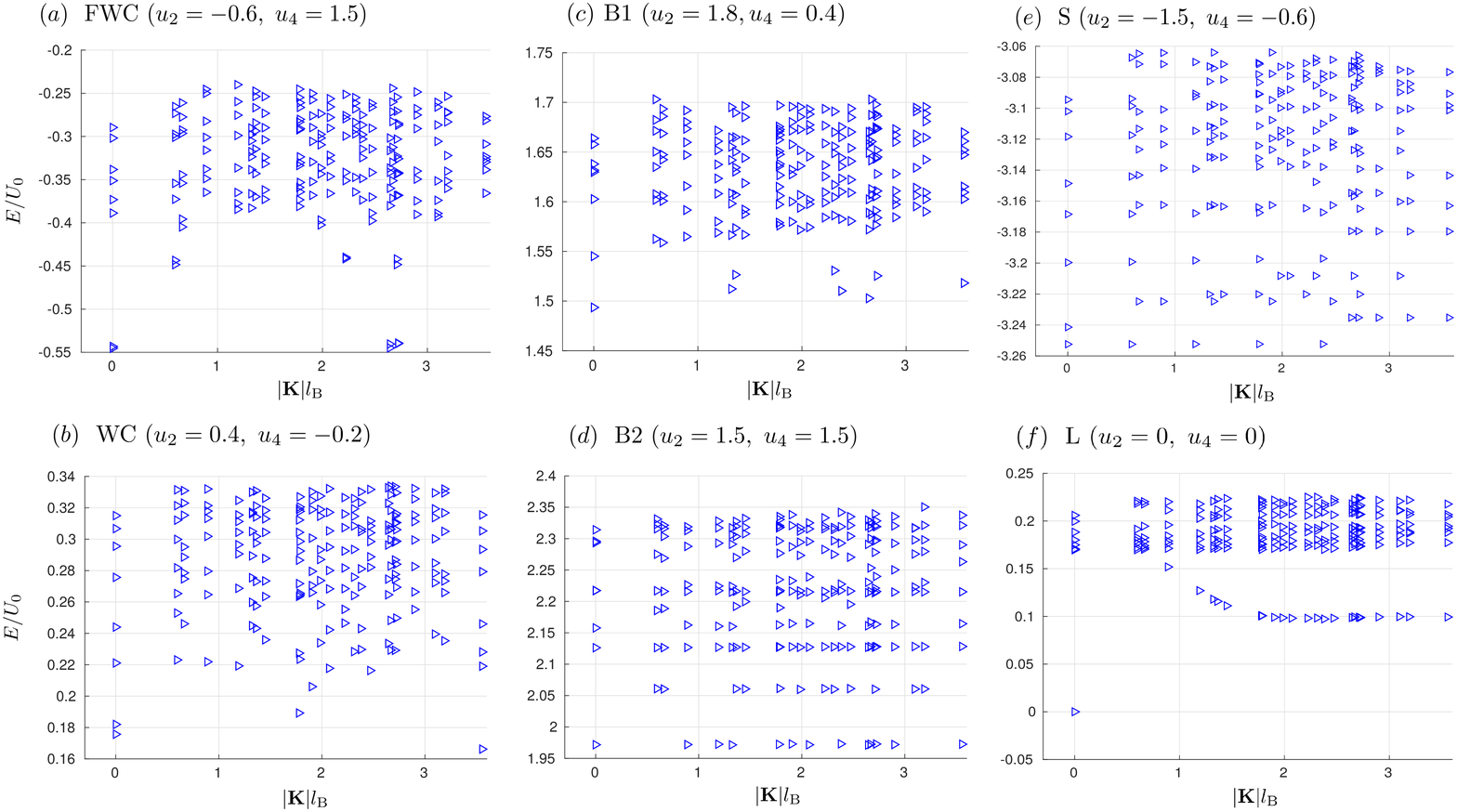}
\caption{\label{spectra} For the same parameters $u_2, u_4$ as used in Fig. 2 of the main text, we plot the energy spectra ($N=8, a/b=0.9$). Pseudomomenta $|{\bf K}|$ are in units of the inverse magnetic length $l_{\rm B}\equiv\sqrt{2\pi N_\phi/(ab)}$.}
\end{figure*}

\section{Energy spectra in different phases}

In Fig. 2 of the main text, we plotted ground state pair correlation functions in order to identify the symmetry-broken order. Each phase can also be characterized by its energy spectra which, for the same parameters as in Fig. 2 of the main text, are plotted in Fig. \ref{spectra}.

\end{document}